\documentclass{acm_proc_article-sp}
\usepackage{algorithmic}
\usepackage[ruled,section]{algorithm}
\usepackage{algorithm}
\usepackage{algorithmic}
\usepackage{epsfig}
\usepackage{amssymb}
\usepackage{amsmath}
\usepackage{amsfonts}
\usepackage{graphicx}

\newcommand{\comment}[1]{}

\begin{document}

\title{Efficient Query Verification on Outsourced Data: A Game-Theoretic Approach}

\author{Robert Nix and Murat Kantarcioglu\\
				The University of Texas at Dallas\\
				500 W Campbell Rd\\
				Richardson, TX 75080\\
				\{rcn062000,muratk\}@utdallas.edu}
\maketitle

\begin{abstract}
To save time and money, businesses and individuals have begun outsourcing their data and computations to cloud computing services.  These entities would, however, like to ensure that the queries they request from the cloud services are being computed correctly.  In this paper, we use the principles of economics and competition to vastly reduce the complexity of query verification on outsourced data.  We consider two cases: First, we consider the scenario where multiple non-colluding data outsourcing services exist, and then we consider the case where only a single outsourcing service exists.  Using a game theoretic model, we show that given the proper incentive structure, we can effectively deter dishonest behavior on the part of the data outsourcing services with very few computational and monetary resources.  We prove that the incentive for an outsourcing service to cheat can be reduced to zero.  Finally, we show that a simple verification method can achieve this reduction through extensive experimental evaluation.
\end{abstract}


\section{Introduction}

As the amount of data that we generate increases, so does the time and effort necessary to process and store the data.  With an increase in time and effort comes an increase in monetary cost.  To this end, many have turned to outsourcing their data processing to ``the cloud.''  Cloud computing services are offered by many large companies, such as Amazon, IBM, Microsoft, and Google, as well as smaller companies such as Joyent and CSC.  For example, Google \cite{bigquery} recently launched the Google BigQuery Service, which is designed for exactly this purpose: outsourced data processing.  The distributed nature of these cloud services shortens data processing time significantly, and offers a massive amount of storage.  

In a perfect world, these cloud providers would impartially devote all the computation necessary to any task paid for by the subscribers.  In such a world, the querying process would look like figure 1 (minus the verifier), where the subscriber outsources the data $D$ to the cloud, and sends queries ($Q$), and the cloud does the necessary calculations and returns the result ($Q(D)$).  However, a cloud provider is a self-interested entity.  Since it is very difficult for the users of the cloud to see the inner workings of the cloud service, a cloud provider could ``cut corners,'' delivering a less accurate or incomplete computation result which would take fewer system resources to compute.  This would, of course, save computational resources for the provider, provided the subscriber was unable to tell a false result from a true one.  Because of this, query verification, or the assurance of query result correctness, has been identified as one of the major problems in data outsourcing \cite{sion2007secure}.

\begin{figure}
\begin{center}
\includegraphics[width=0.8\columnwidth]{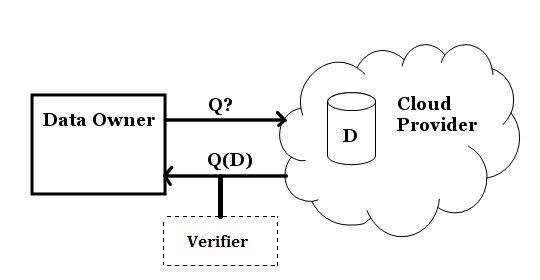}
\caption{Data Outsourcing with Verification}
\end{center}
\end{figure}

Many techniques have been developed and employed for query verification.  Query verification is the process of verifying the authenticity of an outsourced query result.  In figure 1 above, the subscriber sends a query to the outsourcing service, and receives a response.  Query verification would then be another process where the subscriber determines if the response is, in fact, the result of the query.  The verification process may belong to the owner, or it may be another process entirely.  In any case, the verifier aims to make sure that the outsourced server responded correctly.  These verification techniques range from simple to extremely complex, and generally rely on the subscriber storing some sketch of the data (much smaller in size), or some cryptographic protocols.  Such protocols do a good job verifying the data, but are often slow, or only work with specific types of queries.  Many of them require that the subscriber know which queries he will execute in advance, so that a sketch can be created for each one.  None of these, however, consider the heart of the problem: the self-interest of the parties.

The problem of data outsourcing, and the resultant query verification, is fundamentally a problem of \textit{incentives}.  A cloud subscriber wants to get the result of his queries accurately and efficiently, with as low a cost as possible.  A cloud provider, however, is most concerned about the profitable use of its computing resources.  These incentives can be at odds with each other.  The natural way of analyzing competing incentives is to use \textit{game theory}.

Game theory is a branch of economics which studies competitive behavior between parties.  An interaction between parties is cast as a game, where players use strategy to maximize their gains.  The gains from an interaction can be offset artificially by contracts, enforced by law.  These adjustments can make actions which were once profitable, i.e., ``cutting corners'' in a calculation, less profitable through the use of penalties.  The contracts, therefore, aim not to detect whether a cloud provider is cheating, but to remove the incentive for the provider to cheat altogether.  

We propose a game theory-based approach to query verification on outsourced data.  We model the process of querying outsourced data as a game, with contracts used to enforce behavior.  Data outsourcing does not take place in a vacuum.  Service Level Agreements (SLAs) exist for all types of cloud services\cite{patel2009service}, and are enforceable contracts in court.  Thus, we can augment the SLA with an incentive structure to encourage honest behavior.  Using a very simple query verification technique, we show that even the threat of verification is enough to deter cheating by a cloud provider.

First, we consider the case where multiple, non-colluding cloud providers exist.  Non-colluding means that the cloud providers do not share information.  We feel this is realistic, since cloud providers are competing entities and do not wish to share data with the competition.  In this scenario, we show that without the use of special verification techniques, a data owner can guarantee correct results from rational cloud providers, while incurring an additional cost that is only a small fraction of the overall computation cost.

We also consider the case where only a single cloud provider is used.  A data owner may wish to use only a single cloud provider to save money, as they may not have the money to hire multiple cloud services.  In addition, a data owner may simply wish to minimize the outside exposure of the data.  We choose to demonstrate our approach using the simple random sampling query verification technique.  This technique was rejected in many works before, because it required a relatively large storage overhead to achieve a close bound on the sample result.  For our approach, we do not need incredibly close bounds on the result. We only need bounds close enough to catch some mistakes.  The simple random sampling technique also has the added bonus that it can be used to verify many different types of queries, including any aggregates which can be estimated from a sample (count, sum, average, standard deviation, median, quantiles, max, min, etc), and also estimate the size of selection queries, allowing for some verification on those queries as well.  Finally, our method requires running the verification on only a fraction of the queries, incurring a much lower expected runtime than a full sample-based verification method.

Our contributions can be summarized as follows:

\begin{itemize}

\item We develop a game theoretic model of query verification on outsourced data.

\item For both the multiple non-colluding cloud case and the single cloud case, we show that the model has an equilibrium where the cloud provider behaves honestly.

\item We show that a simple sampling technique, although rejected by other works, becomes practical in our single cloud setting.

\item Through extensive experimentation, we show that use of this simple sampling method, coupled with our incentive structure, deters cheating in practice. 

\item Finally, we show that our incentives can improve the expected runtime of \textit{any} query verification method, making it extremely flexible.
\end{itemize}

Our paper does not consider the privacy of the outsourced data (similar to \cite{canetti2011practical}).  However, any privacy-preserving technique for outsourcing data could still be used in our framework.  The use of our game theoretic techniques will not affect the privacy-preserving properties of such schemes.

\section{Related Work}

Several scholarly works have outlined query verification methods.  The vast majority of these works focus on specific types of queries.  Some focus only on selection  \cite{atallah2008efficient,chen2008access,mykletun2006authentication,xie2007integrity,yang2009authenticated}, while others focus on relational queries such as selection, projection, and joins \cite{pang2009scalable,pang2005verifying}.  Still others focus only on aggregation queries like sum, count, and average \cite{haber2006privacy,xu2010authenticating,yi2009small}.  Some of these processes \cite{sion2005query,yi2009small} require different verification schemes for each type of query, or even each individual query, requiring that the subscriber knows which queries will be asked in advance.

Many of the schemes require complex cryptographic protocols.  Some encrypt the data itself, relying on homomorphic schemes to allow the cloud provider to perform the computation \cite{gennaro2010non,xu2010authenticating}.  A homomorphic operation will always be less efficient than the operation on the plaintext, rendering the overhead of these protocols greater by orders of magnitude.  Others, such as \cite{sion2005query}, rely on relatively simpler cryptographic primitives, like secure hash functions.  To maintain integrity, our scheme will also use hash functions.  Our verification framework is, however, simpler than these, and can be used to improve the expected runtime of any of these verification schemes.

The work of Canetti, Riva, and Rothblum \cite{canetti2011practical} also makes use of multiple outsourcing services for query verification.  However, they make use of all the services all the time, and require a logarithmic number of rounds to ensure verifiability of computation.  In addition, they assume that at least one of the cloud providers is in fact honest.  We, in contrast, do not assume that any provider is honest, merely that they are \textit{rational} (meaning that the provider wishes to maximize his profits), and we only use additional providers a fraction of the time.  In addition, we only require one round of computation.
\section{Background}

Before delving into the depths of outsourced query verification, some background knowledge is necessary.  We will require some basic knowledge of game theory.  In addition, we will be making use of some basic cryptographic primitives to ensure data integrity.  

\subsection{Game Theory}

Game theory, despite the misleading name, is a widely accepted field of economic theory which studies competitive behavior.  Competing parties are known as \textit{players}, and the competition itself is known as a \textit{game}.  A game contains four basic elements: \textit{players}, \textit{actions}, \textit{payoffs}, and \textit{information} \cite{rasmusen2007games}.  Players have actions which they can perform at designated times in the game, and as a result of the actions in the game, players receive payoffs.  The payoffs are represented as real-valued functions which depend on the actions chosen and the information surrounding the game.  The players can have different pieces of information, which can have a tremendous impact on the outcome of the game.  The players aim to use a profitable \textit{strategy} to increase their payout.  A player who acts in such a way as to maximize his or her payout, regardless of the effects on other players, is called \textit{rational}.  Games take many forms, and vary in the four attributes mentioned above, but all games deal with them.  The specific games we describe in this paper are finite player, two-step, incomplete information Bayesian games, with payouts based on the final result of players' actions.

A game is said to be at \textit{equilibrium} when no single player can unilaterally increase his or her payoff by changing his or her strategy.  In such a scenario, no players have any incentive to choose a different strategy.  It was shown by Nash \cite{JohnFreakingNash} that all finite player games have an equilibrium, although the equilibrium might require \textit{mixed strategies}.  A mixed strategy is a strategy in which players choose each of the available actions with a certain probability.  For example, consider the game with two players, A and B.  During the game, the players can choose either action X or action Y, and both players choose their actions simultaneously.  If both players choose the same action, player A recieves a utility of 1, while player B recieves a utility of 0.  Otherwise, player B recieves a utility of 1, and player A recieves a utility of 0.  This game can be represented by the table in figure 1.

\begin{figure}
\begin{center}
\begin{tabular}{c|c|c|}
$\frac{A\rightarrow}{B\downarrow}$&X&Y\\
\hline
X&1,0&0,1\\
\hline
X&0,1&1,0\\
\hline
\end{tabular}
\end{center}
\caption{A simple game with a mixed strategy equilibrium}
\end{figure}

Suppose player A's strategy is to always choose action X.  Player B could then choose his action to be Y, and guarantee himself a payout of 1.  However, if this was the case, then player A could simply alter his strategy to choose action Y, thwarting player B's strategy.  Suppose, however, that player A's strategy is to flip a fair coin, choose X if it comes up heads, and tails if it comes up Y.  In this scenario, no matter what player B chooses, player B's expected payout is $\frac{1}{2}$.  Player B can also choose to use this strategy.  If both players use this strategy, then the game is in equlibrium, since neither player has any incentive to unilaterally change strategy.  This equilibrium is the only equilibrium of the game, and since the strategies are probabilistic, the equilibrium is a \textit{mixed strategy} equilibrium.

We can also frame the above game as a game with a pure strategy equilibrium, but with continuous actions.  Instead of having the actions be $X$ and $Y$, we allow each player to select, as his action, a probability between zero and one that they would select $X$.  Let the probability that $A$ chooses $X$ be $\alpha$, and the probability that $B$ chooses $X$ be $\beta$.  As before, the equilibrium is $\alpha = \beta = \frac{1}{2}$.  However, this equilibrium is in pure strategies, since the action is now to choose the probability, not the action as before.  This could be considered an irrelevant distinction.  However, it will prove to be useful in our game theoretic model.

For behavior at an equilibrium to be considered rational, it must not only be \textit{incentive compatible}, meaning that no player has any incentive to unilaterally deviate from that strategy, but it must also be \textit{individually rational}.  Individual rationality means that each player is expected to be no worse off than they were before they chose to participate in the game.  More formally, it means that the payoffs for each player in the equilibrium have an expected value greater than or equal to zero.

\subsection{Cryptographic Primitives}

In order to maintain the integrity of our databases, we will need to employ some basic cryptographic primitives.  We will need to employ a scheme that allows the owner to make sure that tuples he receives from the server are legitimate, and were not added or altered by the server.  We can use a simple message authentication code protocol known as HMAC to do this.  HMAC requires the use of cryptographic hash functions.

A \textit{cryptographic hash function} or \textit{one-way hash function} is a function mapping a large, potentially infinite, domain to a finite range.  This function is simple to compute (taking polynomial time), but is difficult to invert.  Equivalently, we can say that, for a cryptographic hash function $f$, it is difficult to find an $x$ and $y$ such that $x \neq y$ and $f(x) = f(y)$.  Examples of cryptographic hash functions include MD5, SHA-1, and SHA-256.

The HMAC (Hashing Message Authentication Code) system creates a \textit{keyed} hash function from an existing cryptographic hash function.  Let $m$ be the message we wish to create a code for, and $k$ be the key we wish to use.  Let $f$ be our cryptographic hash function, and let its required input size be $n$.  If $k$ has a length smaller than $n$, we pad $k$ with zeroes until it has size $n$.  If $k$ is larger, we let $k$ be $f(k)$ for the purposes of calculating the HMAC function.  We define the HMAC function as follows:
$$HMAC(m,k) = f((k \oplus outpad) || f(k \oplus inpad) || m)$$

Where outpad and inpad are two constants which are the length of $f$'s block size (in practice, 0x5c...5c and 0x36...36, respectively).

Given a message $m$ and its HMAC value $h$, if we have the key $k$, we can simply check to see if HMAC($m$,$k$) matches $h$.  If it does, then the probability that the message is not legitimate (i.e., fabricated or altered), is negligible.

\section{The Multiple Cloud Case}

We first consider the case where multiple cloud providers exist, which do not collude.  This means that the parties do not exchange strategies, and do not exchange information.  We model the query verification process as a game.  The game has the following characteristics:

\textit{Players}: Three players, the Data Owner($O$), and two outsourced servers ($S_1$ and $S_2$).

\textit{Actions}: The data owner begins the game by selecting a probability $\alpha$, and declares this probability to the servers.  He then sends the query ($Q$) to one of the two servers, with equal probability.  With probability $\alpha$ he also sends the query to the other server.  If server $S_i$ receives the query, they then respond to the query with either $Q(D)$, that is, the query result on the database $D$, or $Q'_i(D)$ which is some result other than $Q(D)$. We apply the subscript $i$ to $Q'$ to indicate that one player's method of cheating is different from the other players' method of cheating.  We denote the honest action as $h$, and the cheating action as $c$.  These actions are depicted in figure 3.

\textit{Information}: Data Owner $O$ has given his database $D$ to $S_1$ and $S_2$, with an HMAC message authentication code appended to each tuple.  Any message authentication scheme would work here,  but its purpose and only effect is that it maintains the integrity of the data.  This means that the servers cannot alter any tuples and cannot add any tuples without being detected.  The players have entered into an agreement (a contract) before the game, and the contents of this contract are known to all players.  The contract could contain the probability $\alpha$.

\textit{Payoffs}: The owner recieves the information value of the results received, given by $I_v(Q)$, where $Q$ is either $Q(D)$ or $Q'_i(D)$, minus the amount paid to the servers $P(Q)$.  The servers recieve this payment, minus the cost of computing the query, $C(Q)$. For simplicity's sake, we assume that both outsourcing services have the same cost of computation and receive the same payment for the query.  The logic below easily applies to the case where costs are different, but this assumption simplifies the equations involved.  These payoffs are additionally adjusted by the aforementioned contract.  

We assume that $I_v(Q(D)) \geq (1+\alpha) P(Q)$ and $P(Q) \geq C(Q)$.  If this were not the case, then the game would not be individually rational without some outside subsidies (that is, some player's expected payout would be less than zero).  In essence, we want to ensure that the subscriber would want to pay $(1+\alpha) P(Q)$ to receive the result, and the cloud provider would accept $P(Q)$ for the computation.  To do this, we make sure that the value that the subscriber places on the query is at least the expected payment, and the cost to the cloud providers is no more than the amount they would be paid.  No one takes a loss on the transaction.

\begin{figure}
\begin{center}
\includegraphics[width=0.9\columnwidth]{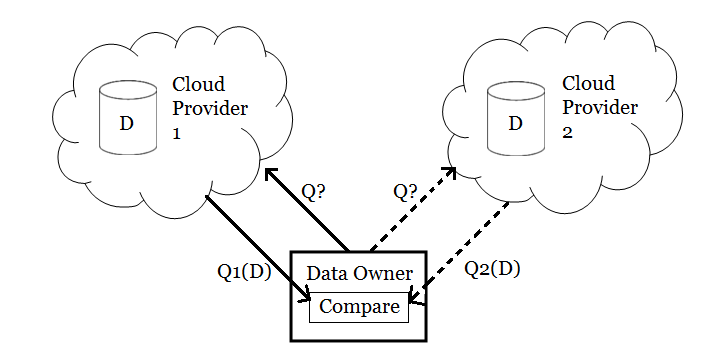}
\end{center}
\caption{The Two-Cloud Query Verification System}
\end{figure}
We now present two contracts, both of which provide simple solutions to the above game in which neither server has incentive to cheat.  The first is very simple and requires no additional computation.  The second is intuitively more fair, and thus more likely to be accepted in a real world scenario.  Both contracts, however, would be accepted by rational players.

\textbf{Contract 1} If the owner asks for query responses from both servers, and the results do not match, both servers pay a penalty of $F$ to the owner, and return the money paid for the computation $P(Q)$ as well.

\textbf{Theorem 4.1} The above game with contract 1 has an individually rational, incentive compatible equilibrium in which the servers behave honestly.

\textit{Proof}: Let $C(Q'_i)$ be the cost of computing $Q'_i$ for $S_i$.  Note that, because $S_1$ and $S_2$ do not collude, $S_1$ does not know $Q'_2$, and $S_2$ does not know $Q'_1$.  The only function both know for sure is $Q$.  Without additional knowledge, we can assume that the probability that $Q'_1(D) = Q'_2(D)$ is negligible.  For a player to even consider returning $Q'_i$ instead of $Q$, we must have $C(Q'_i) \leq C(Q)$, since a player will not cheat if they do not gain anything from it.  We also assume that $I_v(Q'_i(D)) < 0 < I_v(Q(D))$, since not only is the false result not what the owner asked for, but also appears to be the true result if not verified.  If the wrong answer is believed to be correct, this would lead to wrong decisions, and ultimately, financial loss, on the part of the owner.  Now, we can define the expected payoffs to each player, where $u_{P}(x,y)$ is the expected utility for player $P$ when $S_1$ takes action $x$ and $S_2$ takes action $y$.  Note that, in these equations and throughout the rest of the paper, we omit the argument $D$ from $Q$, since $D$ is fixed.  We begin with $O$.  If both players are honest (equation \ref{mhh}), $O$ recieves the value of the information gained from the query, minus the expected payment for the calculation, $1+\alpha$ times $P(Q)$.  If one player is dishonest (equations \ref{mhc} and \ref{mch}), then with probability $\alpha$, $O$ detects this and gets both the honest and the dishonest result and the fine $F$ from both players.  With probability $1-\alpha$, he does not detect this, and gets either the correct value or the incorrect value with equal probability.  In the event that both players cheat (equation \ref{mcc}), they are once again caught with probability $\alpha$, but in this case, when they are not caught, $O$ receives only bogus values.  This results in the following equations:
\begin{align}
u_O(h,h) ={}& I_v(Q(D))-(1+\alpha)P(Q)\label{mhh}\\
u_O(h,c) ={}& \alpha (2F+I_v(Q)+I_v(Q'_2))\label{mhc}\\
				 & + (1-\alpha)(\frac{1}{2}(I_v(Q) + I_v(Q'_2))-P(Q)) \nonumber\\
u_O(c,h) ={}& \alpha (2F+I_v(Q)+I_v(Q'_1))\label{mch}\\
				 & + (1-\alpha)(\frac{1}{2}(I_v(Q) + I_v(Q'_1))-P(Q))\nonumber\\
u_O(c,c) ={}& \alpha (2F+I_v(Q'_1)+I_v(Q'_2))\label{mcc}\\
				 & + (1-\alpha)(\frac{1}{2}(I_v(Q'_1) + I_v(Q'_2))-P(Q)) \nonumber
\end{align}

For the servers, if both servers are honest (equations \ref{ms1hh} and \ref{ms2hh}), they receive the payment for the query, minus the cost of the query, provided they are selected to perform the calculation.  This selection probability is why the equations below contain $\frac{1}{2}$.  Otherwise, they gain nothing and lose nothing.  If one player is dishonest, that player (equations \ref{ms1ch} and \ref{ms2hc}), regardless of whether the other player is honest, with probability $\alpha$ is caught, and loses the fine $F$.  With probability $1-\alpha$, the player is not caught, and gains the payment $P(Q)$, minus the cost of computing his cheat, $C(Q'_i)$, if he is chosen for the computation.  If a player is honest while the other player is dishonest (equations \ref{ms1hc} and \ref{ms2ch}), that player similarly is punished with probability $\alpha$, but invests a cost of $C(Q)$ instead of $C(Q'_i)$ in the computation.  This gives us the following equations:

\begin{align}
u_{S_1}(h,h) &= \frac{1}{2}(1+\alpha)(P(Q) - C(Q))\label{ms1hh}\\
u_{S_1}(h,c) &= \frac{1}{2}(1-\alpha)(P(Q) - C(Q)) - \alpha F\label{ms1hc}\\
u_{S_1}(c,h) &= u_{S_1}(c,c) = \frac{1}{2}(1-\alpha)(P(Q) - C(Q'_1)) - \alpha F\label{ms1ch}
\end{align}

\begin{align}
u_{S_2}(h,h) &= \frac{1}{2}(1+\alpha)(P(Q) - C(Q))\label{ms2hh}\\
u_{S_2}(c,h) &= \frac{1}{2}(1-\alpha)(P(Q) - C(Q)) - \alpha F\label{ms2ch}\\
u_{S_2}(h,c) &= u_{S_1}(c,c) = \frac{1}{2}(1-\alpha)(P(Q) - C(Q'_2)) - \alpha F\label{ms2hc}
\end{align}

We can now find the $\alpha$ for which the expected value for $S_1$ is less when he cheats than when he is honest, assuming $S_2$ is honest.  By symmetry, this will be the same for $S_2$.  Thus, we set:
$$\frac{1}{2}(1-\alpha)(P(Q) - C(Q'_1)) - \alpha F \leq \frac{1}{2}(1+\alpha)(P(Q) - C(Q))$$

Let $H$ represent the quantity $P(Q) - C(Q)$, and $H'$ represent the quantity $P(Q) - C(Q'_1)$. Distribute the $(1+\alpha)$ and $(1-\alpha)$ to get:
\begin{eqnarray}
 \frac{1}{2}(H')-\frac{\alpha}{2}(H')-\alpha F&\leq&\frac{1}{2}(H)+\frac{\alpha}{2}(H) \nonumber
\end{eqnarray}

Rearranging and combining terms, we get:
\begin{eqnarray}
\frac{1}{2}(C(Q) - C(Q'_1)) &\leq& \alpha F + \alpha P(Q)\nonumber\\
& &+ \frac{\alpha}{2} (C(Q) - C(Q'_1))\nonumber
\end{eqnarray}

Let $G$ represent the quantity $C(Q)-C(Q'_1)$, that is, the amount the server would gain from cheating.  Substituting this in and factoring out an $\alpha$, we get:
$$\frac{1}{2}G \leq \alpha (F+P(Q)+\frac{1}{2}G)$$

Multiplying through by two, we get:
$$G \leq \alpha (2F+2P(Q)+G)$$

And, solving for $\alpha$,

\begin{align}
\frac{G}{2F+2P(Q) + G} &\leq \alpha  \label{solvedalpha}
\end{align}

Since we can define $F$ to be whatever we want in the contract, we can make this minimum $\alpha$ value arbitrarily small.  If $\alpha$ is at least this much, then $S_1$ (and by symmetry, $S_2$) has no incentive to cheat.  If $S_2$ is not honest, then $S_1$ has no incentive to be honest, but the payout is less for both (much less, if $F$ is large).  Therefore, the best outcome is for both players to behave honestly.    

Now, we need to show that choosing $\alpha$ is incentive compatible for $O$.  Given that both players are honest, $O$'s utility is given as:
$$u_O(h,h) = I_v(Q(D)) - (1+\alpha)P(Q)$$

Which, by our assumption, is greater than or equal to zero.  Thus, it is individually rational for $O$.  If $\alpha$ is increased, it merely decreases this value, so $O$ has no incentive to increase $\alpha$.  If we decrease $\alpha$, then $S_1$ and $S_2$ will start cheating!  This leads to:

\begin{align*}
u_O(c,c) ={}& \alpha (2F+I_v(Q'_1)+I_v(Q'_2))\\
				 & + (1-\alpha)(\frac{1}{2}(I_v(Q'_1) + I_v(Q'_2))-P(Q))
\end{align*}

Now, since our $\alpha$ is less than our prescribed value in equation (\ref{solvedalpha}), $F$ is bounded above by $\frac{G}{\alpha}-2P(Q)-G$.  Because of this, as $\alpha$ tends to zero, the first term of the above equation decreases.  The second term is negative (as $I_v(Q'_1)$ and $I_v(Q'_2)$ are less than zero), and gets worse as $\alpha$ tends to zero.  Thus, if $\alpha$ is less than our prescribed value, $O$ expects to lose value from cheating.   So, $O$ has no incentive to deviate from $\alpha = \frac{G}{2F+2P(Q)+G}$.

Now, in practice, $O$ does not know $G$.  Thus, he must choose the smallest $\alpha$ that he knows he can use.  Since $P(Q) \geq C(Q) \geq G$, $O$ can choose $\alpha = \frac{P(Q)}{2F+2P(Q)-P(Q)} = \frac{P(Q)}{2F-P(Q)}$.  

As this is both incentive compatible and individually rational for all players, this contract creates the best possible equilibrium where $S_1$ and $S_2$ do not cheat, and $O$ pays only $(1+\alpha)$ times the price of a single computation (where $\alpha$ is small).\qed

Now, it might seem unfair to punish both players when only one player cheats.  The rational player would see the above contract as completely fair, but humans are not always completely rational.  Thus, we also briefly examine a contract which identifies the cheater and punishes only the cheater.

\textbf{Contract 2} If the owner asks for query responses from both servers, and the results do not match, the owner performs a potentially costly audit of the computation.  Each server whose result does not match the result given by the owner's process pays a fine $F$ to the owner.

The audit process mentioned above could be done in several ways.  The simplest, although most expensive, of these would be for the owner to retrieve all the data, then perform the query himself.  Obviously, this defeats the purpose of data outsourcing.  Based on the fact that the outsourced data uses some message authentication codes to keep the data from being modified, we can improve this.  First, for selection queries, if one player fails any MAC checks, then they are obviously cheating.  If one player returns fewer results than the other, then they are also obviously cheating.  For aggregate queries, we can have each source return the tuples which were selected for the aggregation process.  We can then check to see if the aggregate query result matches the values returned by the server.  Finding a tuple set that matches a false query result might prove incredibly difficult, if the false query was not generated from a sample.  We can also apply the same techniques used for selection queries, noting that the cloud that returns fewer tuples must be cheating (provided all tuples returned are authenticated).  Essentially, for a given query, we end up asking the clouds to ``show their work,'' or face consequences.

\textbf{Theorem 4.2} The above game under contract 2 also has an equilibrium where both servers remain honest.

\textit{Proof}:  The main differences between this and the previous scenario are the fact that an honest server will never pay a fine, and that if a player is caught cheating, the owner must perform a costly audit.  We will call this cost $C(Q_O(D))$.  As the data is signed with the HMAC codes, the owner can retrieve all of it from either server, guaranteed.  As $u_{S_1}(h,h)$ and $u_{S_1}(c,h)$ do not change, the $\alpha$ may be located in the same way.  When one player cheats, the other player has incentive to be honest, as it avoids the fine.  Thus, $(h,h)$ is actually a \textit{dominant strategy} in this game, when $\alpha$ is set high enough.  Now, recall that $F$ can be set as high as necessary.  If we double $F$ and increase it by the cost of the audit $C(Q_O(D))$, then the payouts for $O$ would be the same as in contract 1.  So, by Theorem 4.1, this is also incentive compatible for $O$.

As this is both incentive compatible and individually rational for all players, this contract creates the best possible equilibrium where $S_1$ and $S_2$ do not cheat, and $O$ pays only $(1+\alpha)$ times the price of a single computation (where $\alpha$ is small).  Note that, since both $S_1$ and $S_2$ are honest, we \textbf{never} expect to pay the cost of the audit.\qed

Note the generality of this result.  In contrast with many other results, it works for \textbf{any query} on \textbf{any database} (with the caveat that the query is deterministic), and it works in only one round of computation.

\section{The Single Cloud Case}

Though the above scenario is quite simple and very efficient, it does require giving both money and data to multiple parties.  It might be that the cost of maintaining two cloud services (due to storage costs and other overhead) is prohibitively expensive.  A data owner might also want to minimize the outside exposure of his data set.  It is possible, then, to use a similar scheme to verify the result from a single cloud.  For the single cloud case, we focus on a database with a single relation.  The extension to include joins will be considered in future work.  We once again cast the process of query verification as a game.  The game has the following characteristics:

\textit{Players}: Two players, the Data Owner ($O$) and the outsourced Server($S$).

\textit{Actions}: The data owner begins the game by selecting a probability $\alpha$, and declares this probability to the server.  This probability $\alpha$ is the probability with which the result of the query ($Q$) will be verified ($v$).  With probability $1-\alpha$, the query will not be verified ($n$).  After receiving this probability, the server may choose to cheat ($c$), revealing $q'_S = Q'(R)$, an incorrect result, or honestly ($h$) give the result $q_S = Q(R)$.  The data owner then verifies with the probability $\alpha$, first by performing a local evaluation, then, if necessary, a full query audit.

\begin{figure*}
\begin{center}
\includegraphics*[width=0.7\linewidth]{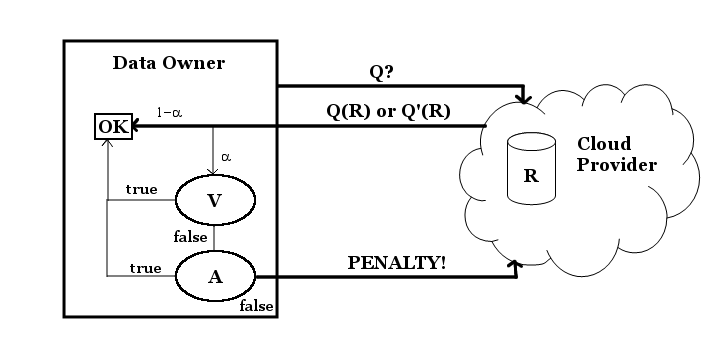}
\end{center}
\caption{Verifying Queries With a Single Cloud}
\end{figure*}

\textit{Information}: $O$ has given his database relation ($R$) to $S$, along with authenication codes for each tuple (to prevent modification).  $O$ retains a sketch of the database ($R'$) which will be used for verification.  $O$ has a process $V(Q,q)$ which determines whether the argument $q$ is equal to $q_S$ with high probability, using the sketch $R'$.  In addition, an auditing method exists $A(Q,q,R)$ which will determine with certainty whether the query was executed correctly, but is very expensive.  The players have entered into a contract before the game, and the contents of the contract are known to all players.  \comment{which states that if a query is successfully executed (i.e., $O$ chooses not to verify, $O$ incorrectly verifies a cheat, or $S$ chooses to give the correct result), $S$ receives some amount of money ($P(Q)$, as this amount may vary based on the query) from $O$ for his services.  If, however, $S$ is caught cheating, he pays a fine ($F$) to $O$.  If the verification process requires auditing, and $S$ was not cheating, then $O$ covers the cost of the auditing for $S$.}  This contract can adjust the payoffs below.

\textit{Payoffs}: Let $p_{tp}$ be the probability that $V(Q,q_S)$ returns true, $p_{tn}$ be the probability that $V(Q,q'_S)$ returns false, $p_{fp} = 1 - p_{tn}$, and $p_{fn} = 1 - p_{tp}$ (These are the probabilities of true positives, true negatives, false positives and false negatives from $V$, respectively).  Let $C(X)$ be the cost of computing the expression $X$.  Let $I_v(X)$ be the value of the information given by $X$.  The expected utilities (payoffs) for each player, without the intervening contract, are as follows:

When the owner decides not to verify, he simply receives the value of the query result (honest or not), minus the amount paid for the calculation, resulting in:
\begin{align*}
u_{O}(n,h)&=I_v(q_S) - P(Q)\\
u_{O}(n,c)&=I_v(q'_S) - P(Q)
\end{align*}

Similarly, the server simply gains the amount paid, minus the cost of the calculation:
\begin{align*}
u_{S}(n,h)&=P(Q) - C(q_S)\\
u_{S}(n,c)&=P(Q) - C(q'_S)
\end{align*}

If the owner chooses to verify, he also pays the cost of verification, and in the case of a failed $V$, also pays the cost of an audit.  If the audit fails (which would only happen in the case of a cheating server), he does not pay the price for the calculation.
\begin{align*}
u_{O}(v,h)={}&I_v(q_S) - P(Q) - C(V(Q,q_S))\\
  &  - p_{fn}\cdot C(A(Q,q_S,R))\\
u_{O}(v,c)={}&I_v(Q'(R)) - C(V(Q,q'_S))\\
  & - p_{tn}\cdot C(A(Q,q'_S,R)) - p_{fp}\cdot P(Q)
\end{align*}

An honest server, in the case of the verification, gets the same payout he would without verification.  This is the price of the query minus the cost to calculate it.  A cheating server is only paid if he is not caught, so he is only paid in the case of a false positive from $V$.
\begin{align*}
u_{S}(v,h)&=P(Q) - C(q_S)\\
u_{S}(v,c)&=p_{fp}\cdot P(Q) - C(q'_S)
\end{align*}

Now, since $O$ declares a verification probability in advance, we can write the above as:
\begin{align*}
u_O(\alpha,h) &= \alpha u_O(v,h) + (1-\alpha) u_O(n,h)\\
u_S(\alpha,h) &= \alpha u_S(v,h) + (1-\alpha) u_S(n,h)\\
u_O(\alpha,c) &= \alpha u_O(v,c) + (1-\alpha) u_O(n,c)\\
u_S(\alpha,c) &= \alpha u_S(v,c) + (1-\alpha) u_S(n,c)\\
\end{align*}
Note that, in practice, a payment might not be rendered for every query, and instead the server might charge a flat fee for its services, or some other payment structure.  In these cases, one could consider the total payments spread out throughout the queries.  This assumption that payment is rendered for each query will not invalidate our solution.

We make some assumptions about the values used above.  First, we assume that  $I_v(q_S) \geq P(Q) \geq C(q_S)$.  This is because this inequality is necessary for participation in the game to be individually rational (since this guarantees that the best expected payoff for each player, assuming no one cheats, is at least zero).  Naturally, if the query were not worth enough to the owner, he would not pay the price, and if the price did not cover the cost of computation for the server, he would not perform the calculation.  Second, we assume that as $q'_S$ approaches $q_S$, $C(q'_S)$ approaches $C(q_S)$.  This implies that it is difficult to compute a $q'_S$ such that $V(Q,q'_S)$ is expectedly true.  As $q'_S$ moves away from $q_S$, the cost can decrease.  This provides the initial incentive for the server to cheat.  These assumptions are logical, since computing a value close to actual result becomes more and more like computing the actual result.  For example, if a cheating server were to run the query on a sample of the data and extrapolate the result, the estimated result would get more accurate as the sample size got larger, but the computational resources used would also increase.  We assume that the cost of $V$ and $A$ are constant for a given $Q$ (no matter if $q_S$ or $q'_S$ is provided to them as an argument).  Finally, we once again assume that $I_v(q'_S) < 0 < I_v(q_S)$, due to the result not being what $O$ asked for.  We also assume that $C(A(Q,q'_S,R)) < I_v(q_S) - I_v(q'_S)$, since if the audit were more costly than the amount of information supplied by the query, the audit would not take place. 

We now outline a contract which removes the incentive for the server to cheat.  It is quite similar to the contract for the two-cloud case.

\textbf{Contract 3} If the owner chooses to verify, and it is determined that the server has cheated, the server pays a penalty of $F + C(A(Q,q'_S,R))$.  (Note: We explicitly force a cheating server to pay the audit cost.)

\textbf{Theorem 5} The game, using the above contract (depicted in figure 4), has an equilibrium in pure strategies.  $O$ will select an $\alpha$ which makes cheating less profitable (expectedly) than correctly revealing the result. $S$ chooses not to cheat.

\textit{Proof}: The above contract makes the following changes to the original utilities:

\begin{tabular}{rcl}
$u_{O}(v,c)$&$=$&$I_v(q'_S) - C(V(Q,q'_S))$\\
 & & $- p_{tn}\cdot F - p_{fp}\cdot P(Q)$\\
$u_{S}(v,c)$&$=$&$p_{fp}\cdot P(Q) - C(q'_S)$\\ 
 & & $- p_{tn} \cdot (C(A(Q,q'_S,R))+F)$\\ 
\end{tabular}

Recall that the expected payouts payouts are:

\begin{align*}
u_O(\alpha,h) &= \alpha u_O(v,h) + (1-\alpha) u_O(n,h)\\
u_S(\alpha,h) &= \alpha u_S(v,h) + (1-\alpha) u_S(n,h)\\
u_O(\alpha,c) &= \alpha u_O(v,c) + (1-\alpha) u_O(n,c)\\
u_S(\alpha,c) &= \alpha u_S(v,c) + (1-\alpha) u_S(n,c)\\
\end{align*}

We want to find the $\alpha$ such that $u_S(\alpha,h) \geq u_S(\alpha,c)$.        

Substituting in, we get:     

\begin{tabular}{rcl}
$u_{S}(\alpha,h)$&$=$&$\alpha(P(Q) - C(q_S))$\\
 & &$+(1 - \alpha)(P(Q) - C(q_S))$\\
 &$=$&$P(Q) - C(q_S)$\\
$u_S(\alpha,c)$&$=$&$\alpha(p_{fp}P(Q) - C(q'_S)$\\ 
 & & $- p_{tn}(C(A(Q,q'_S,R))+F))$\\
 & & $+ (1 - \alpha)(P(Q) - C(q'_S))$
\end{tabular}

So, we have the inequality (after multiplying through by $\alpha$ and $1-\alpha$:

\begin{tabular}{rcl}
$P(Q)-C(q_S)$&$\geq$&$\alpha p_{fp}P(Q) - \alpha C(q'_S)$\\ 
 & &$- \alpha p_{tn}(C(A(Q,q'_S,R))+F)$\\
 & &$+ P(Q) - C(q'_S) -\alpha P(Q)$\\
 & &$+ \alpha C(q'_S)$
\end{tabular}

Rearranging terms, we get:

\begin{align*}
P(Q) - P(Q) + C(q'_S) - C(q_S) \geq {} & \alpha(p_{fp}P(Q) - C(q'_S)\\
 & - p_{tn}(C(A(Q,q'_S,R))+F)\\
 & - P(Q) + C(q'_S))
\end{align*}

Canceling out like terms, we get:

\begin{align*}
C(q'_S) - C(q_S) \geq {}& \alpha(p_{fp}-1)P(Q) \\
  & - \alpha p_{tn}(C(A(Q,q'_S,R))+F)
\end{align*}

Now, since $Q'$ is easier to compute than $Q$, $p_{fp}<1$, both sides of this inequality are negative.  We therefore multiply both sides by $-1$ and simplify to get the following:

\begin{align*}
C(q_S) - C(q'_S) \leq {} &\alpha((1-p_{fp})P(Q)\\
 & + p_{tn}(C(A(Q,q'_S,R))+F))
\end{align*}

Since $1-p_{fp}$ is equal to $p_{tn}$, we can substitute in $p_{tn}$, then divide by the coefficient of $\alpha$, yielding the final expression:
$$\alpha \geq \frac{C(q_S)-C(q'_S)}{p_{tn}(C(A(Q,q'_S,R))+F+P(Q))}$$

When $\alpha$ increases, the payout for cheating decreases, provided $C(A(Q,q'_S,R))$ and $F$ are large enough.  So, as long as the expression above is satisfied, the server will choose not to cheat.

Now, while $C(A(Q,q'_S,R))$ is fixed, $F$ is something that can be adjusted in the contract!  Therefore, if the penalty $F$ is astronomically high, we can severely reduce $\alpha$, while maintaining that there is no incentive to cheat for $S$.  This is what is known as a ``boiling-in-oil'' contract \cite{rasmusen2007games}.  

We must also show that this $\alpha$ is incentive compatible for $O$.  Consider what happens when $\alpha$ is increased.  If $\alpha$ is greater than the above value, $O$ ends up verifying more, while $S$ continues telling the truth.  Because of this, $O$ loses valuation.  So, $O$ will not choose $\alpha$ higher than this.  If $\alpha$ is less than this value, then $S$ will start cheating!  The possible increase in payout to $O$ would be $\alpha F$, but since $\alpha$ is small, and $I_v(q_S)$ is so much greater than $I_v(q'_S)$, this would likely result in a decrease in payout for $O$.  Therefore, $\alpha$ is not less than the above expression either.  Thus, we have an equilibrium.\qed

\section{Implementation Details}

The game outlined above is fairly general, and allows for any local verification method $V$ to be used.  Here, we outline a simple sampling verification method which becomes much more viable when the verification process is not being run with every query.  First, let us assume that the data consists of $N$ signed tuples, each of which has a unique, consecutive id from $1$ to $N$.  Let $O$ maintain some sample of size $k$ of these $N$ tuples, together with the value $N$.  This sample is selected uniformly at random from the entire data set, with replacement.  This sample can be used to compute $V(Q,q'_S)$ for many different types of queries.  For aggregate queries such as count, sum, average, standard deviation, etc, one could simply perform the action on the $k$ tuples, and extrapolate based on $N$.  If this sample value is within some $\varepsilon$ of the query result $q'_S$, then we declare the result correct.  Otherwise, we perform the audit.

For selection queries, note that because each tuple is signed, we know that the server cannot modify any tuples, nor can it insert new tuples.  It can only either remove relevant tuples from the result, or insert irrelevant tuples into the result.  If the server inserts irrelevant tuples, this can be easily verified by $O$ by simply noting that the tuple does not match the query.  Thus, it is only difficult to verify when a tuple has been left out.  As before, we can perform the selection query on the sample of $k$ tuples, and extrapolate the number of tuples that should be returned by $q_S$.  If the number of tuples in $q'_S$ is within some $\varepsilon$, we declare the result correct.  If the number of tuples in $q'_S$ is greater than our estimate, then we should also declare the result correct, since a greater number of tuples cannot be wrong.  Otherwise, we perform the audit.

There are plenty of other methods used to verify queries on outsourced data, and \textbf{any} of them would work as a verification method $V$ in our scheme.  We choose this one, however, because of its simplicity.  Note that it does not require expensive cryptographic operations.

One thing remains in the definition of the verification mechanism, and that is the definition of $\varepsilon$.  As the selection of $k$ tuples can be considered a selection of $k$ random variables $X_1, ..., X_k \in R$, and in each case we are interested in a function $f$ which maps $R \rightarrow \Re$, and any alteration in a given $X_i$ can only change the value of the aggregate function by at most some $c_i$ (this $c_i$ is 1 for count, the max value of the given attribute for sum, the max value squared for standard deviation, etc), we can apply McDiarmid's inequality \cite{mcdiarmid}, giving us the following: $$Pr\{|E[f(X_1,...,X_k)]-f(X_1,...,X_k)|\geq \varepsilon\} \leq 2e^{-\frac{2\epsilon^2}{\sum_{i=1}^k c_i^2}}$$

Note, this inequality does not depend on the value of $N$.  It simply depends on the sample size.  For example, say we want to devise a sample size $k$ such that the probability that an average query on attribute $a$ of the sample is within $\varepsilon = 1\%$ of the true result with probability $.999$.  $c_i$ is given as $\frac{\textrm{max}\{|a|\}}{k}$.  The probability in the above works out to $2e^{-\frac{.0002 \textrm{average}\{a\}^2}{\textrm{max}\{|a|\}^2/k}}$.  We want this to be less than or equal to $.001$.  Solving for $k$, we get $$ln(.0005) \leq -\frac{.0002\textrm{average}\{a\}^2 k}{max\{|a|\}^2}$$  $$\frac{-ln(.0005)max\{|a|\}^2}{.0002 \textrm{average}\{a\}^2} \leq k$$

This gives us a $k$ value of approximately $38004.51$ times the maximum value of the attribute $a$, divided by the square of the result.  As the average of the result is no more than the maximum value of $a$, but its square can be much larger, $k$ can be 38 thousand tuples, or less, depending on the distribution of $a$, even if the number of tuples is in the millions.  Note that this is does not help the data owner find the value of $k$, as the owner does not know the actual result.  This merely shows that a good $k$ exists, and it is independent of the number of tuples in the dataset for many common queries.

38,000 tuples is not a particularly small number, especially with some sketches using only three bytes \cite{yi2009small}.  However, this sample can be used to verify many different types of queries, and does not have to plan for the queries in advance.  In addition, the verification will only be performed a fraction ($\alpha$) of the time.  This fraction, through the use of the penalty in the contract, can be made arbitrarily small, leading to a very fast expected runtime.

Now, one method of generating a false query (for aggregate queries) might be to use the same sampling method as above.  Note, however, that in order to ensure that the sample chosen by the server has a result within $\varepsilon$ of the result of the owner's sample, the server would need a much tighter epsilon.

With some probability $\delta$, the owner's sample result is within $\varepsilon$ of the correct result.  This is also true for the server.  However, consider the worst case where the owner's sample value is $Q(R) - \varepsilon$ and the server's sample value is $Q(R) + \varepsilon$.  The probability $\delta$ is not sufficient to bound the difference between the two sample values.  To ensure that with probability $\delta$ the owner's sample result is within $\varepsilon$ of the result returned by the server (with the given high probability), the server would have to return the actual result, as any leeway would lead to a worst case scenario where the difference is greater than $\varepsilon$.

In order to prevent sampling bias, a protocol could be implemented to resample from the data.  As each tuple has a unique id, the owner could, at some interval, request the tuples with a given set of ids.  The owner would know if the tuples he desired were not returned.  In order to prevent the server from learning the exact sample (which would lead to the server simply using that sample for calculations), the owner would select some dummy tuples, or in some cases, the entire data set.  A similar method, selecting all tuples involved, can be employed for auditing the queries.

\section{Experiments}

To test the effectiveness of the sampling protocol for catching cheating on real data, we ran a series of experiments.  The mechanisms outlined in sections 4 and 5 do not need any experimentation, as they are proven and mathematically sound. These experiments were designed to show that the sampling technique can identify cheating with a non-trivial probability.  Other verification methods will work similarly in our framework, as long as they can identify cheating with non-trivial probability.  For example, if a simple method exists to verify a certain query deterministically, then it could be called in place of our sampling scheme, and would allow our $\alpha$ to be even smaller.  The sampling protocol is important, however, due to its generality and simplicity.

\subsection{Methodology}

We used the US Census 1990 data set from the UC Irvine Machine Learning repository, which contains over 2.3 million tuples \cite{uci}.  We focused on a few major fields in this data set.

We processed results for eight different aggregation queries of varying types.  Since selection queries can be estimated via counts, we chose to focus on aggregation queries.

The query types are as follows:

\begin{enumerate}

\item Count, equality selection (count the people whose race value is 2--black)

\item Count, range selection (count the people whose income is greater than 40000)

\item Count, range and equality conjunction (count the people who are over age 30 and never married)

\item Count, range disjunction (count the people who are under age 18 or have an income of less than 10000)

\item Sum, equality selection (find the sum of the incomes of all people who never married)

\item Sum, range and equality conjunction (find the sum of the incomes of all people who are over age 40 and whose place of birth is the place they work)

\item Average, range selection (find the average age of all people who have an income greater than 80000)

\item Average, equality conjunction, sparse result (find the average income of all people who are male and of race 9--Japanese)
\end{enumerate}

For each query, we ran 100 trials, estimating the full result of the query with five different sample sizes: 1000 tuples, 5000 tuples, 10000 tuples, 20000 tuples, and 40000 tuples.  As above, these samples are selected uniformly at random with replacement.  We determined the likelihood that each sample would accept the correct value for varying values of $\varepsilon$ from 0 to $.5r$ where $r$ is the estimated result.  Since the verification process would not know the actual result, we based the $\varepsilon$ on the estimated result given by the sample, as we expected it to be close to the actual result.

We then ran the samples against different means of falsifying the result, to determine if the sample method could catch a cheater.  The first type of falsification was the same as our verification technique, sampling the data.  We once again ran 100 trials for 1000, 5000, 10000, 20000, and 40000 tuple samples, both for detecting the cheating and for creating the cheating.

The second type of falsification was a ``worst case'' falsification, where the exact result was computed, but then Laplace noise was added to the final result.  An adversary would never actually do this, as it would be more expensive than simply computing the result itself, but this provides a way to test our scheme beyond the normal means.  The mean of the Laplace noise was of course the result itself (which we will call $r$), whereas the width parameter was varied from $r/5$, $r/10$, $r/20$, and $r/50$.  We chose Laplace noise as opposed to any other type of noise because it is used in differential privacy as a means of masking query results while still achieving meaningful results \cite{dwork2006differential}.  Each of these sets of noise ran 100 trials against each sample size as before.

\subsection{Results}

Space restrictions prevent the inclusion of every graph generated by the experiments.  However, if we examine one factor at a time, we can show the general trend of the sampling protocol to correctly or incorrectly identify cheating values.  The omitted graphs show similar trends.

\begin{figure*}
\begin{center}
\begin{tabular}{cccc}
\includegraphics[width=0.2\linewidth]{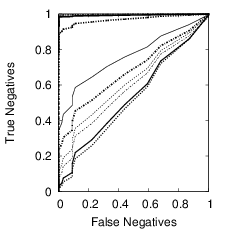}&
\includegraphics[width=0.2\linewidth]{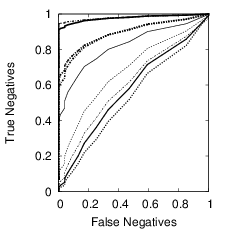}&
\includegraphics[width=0.2\linewidth]{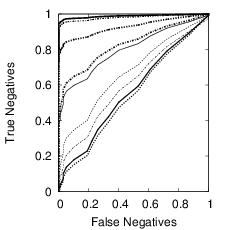}&
\includegraphics[width=0.2\linewidth]{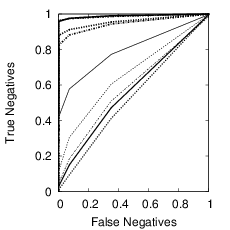}\\
Query 1&Query 2&Query 3&Query 4\\
\includegraphics[width=0.2\linewidth]{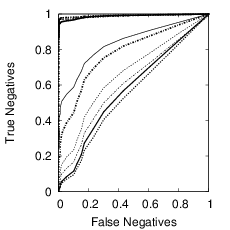}&
\includegraphics[width=0.2\linewidth]{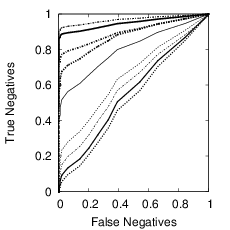}&
\includegraphics[width=0.2\linewidth]{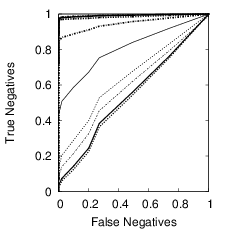}&
\includegraphics[width=0.2\linewidth]{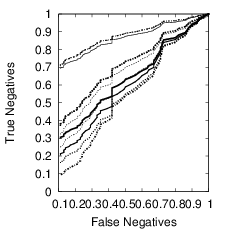}\\
Query 5&Query 6&Query 7&Query 8
\end{tabular}
\includegraphics*[width=0.8\linewidth]{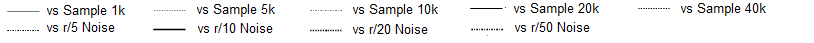}
\end{center}
\caption{ROC Curves for the 8 Query Types: Sample Size 10000}
\end{figure*}

\textbf{Query Type}  Figure 5 shows the ROC (receiver operating characteristic) curves for each of the eight queries, for a sample size of 10000, against every type of result falsification we used.  These ROC curves shows the tradeoff between the probability of a false negative and the probability of a true negative.  The queries themselves all behave similarly.  At a sample size of 10000, we can always find an $\varepsilon$ where some nontrivial fraction of cheating will be caught. There is always a tradeoff, however. As $\varepsilon$ decreases, more legitimate results will be marked as wrong, and forced to be fully audited.  Proper use of the sampling technique involves careful selection of epsilon in order to increase $p_{tn}$, while reducing $p_{fn}$ as much as possible.  In practice, it is more important that $p_{tn}$ be high, since we can mitigate the effect of false positives by increasing the penalty, thereby decreasing $\alpha$ and reducing the number of times that we do the verification.  $p_{tn}$ is acceptably high for fairly small $\epsilon$ ($0.00125r$ to $0.005r$).

\begin{figure*}
\begin{center}
\begin{tabular}{ccc}
\includegraphics[width=0.2\linewidth]{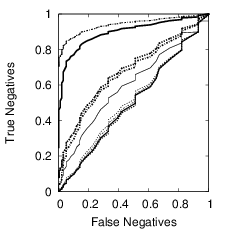}&
\includegraphics[width=0.2\linewidth]{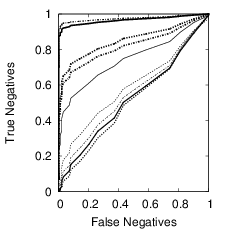}&
\includegraphics[width=0.2\linewidth]{income40kROC10000.png}\\
Sample Size 1000&Sample Size 5000&Sample Size 10000
\end{tabular}
\begin{tabular}{cc}
\includegraphics[width=0.2\linewidth]{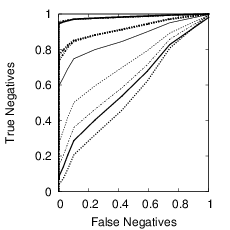}&
\includegraphics[width=0.2\linewidth]{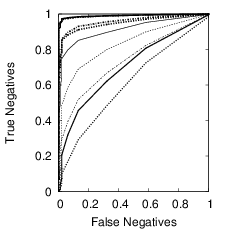}\\
Sample Size 20000&Sample Size 40000
\end{tabular}
\includegraphics*[width=0.8\linewidth]{ROClegend.png}
\end{center}
\caption{ROC Curves for Five Sample Sizes: Query 2}
\end{figure*}

\textbf{Sample Size} Figure 6 shows the ROC curves for query 2 for each sample size used.  Clearly, as the sample size increases, the potential for better choices of epsilon increases.  With a sample size of 40000, there is even a type of cheating ($r$/5 noise) that allows for a false negative rate of .02 and a true negative rate of .95!  The obvious tradeoff here, though, is that while you will do fewer full audits with a larger sample size, the verification process will take more resources.  The smaller sample sizes still have the ability to catch cheating, but they will end up auditing many more legitimate results.

\begin{figure*}
\begin{center}
\begin{tabular}{ccc}
\includegraphics[width=0.2\linewidth]{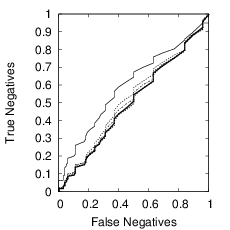}&
\includegraphics[width=0.2\linewidth]{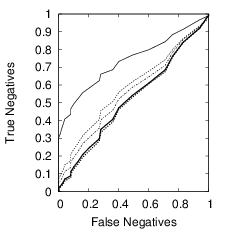}&
\includegraphics[width=0.2\linewidth]{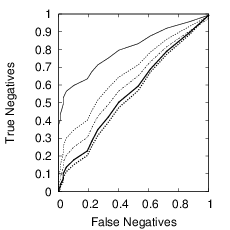}\\
Sample Size 1000&Sample Size 5000&Sample Size 10000
\end{tabular}
\begin{tabular}{cc}
\includegraphics[width=0.2\linewidth]{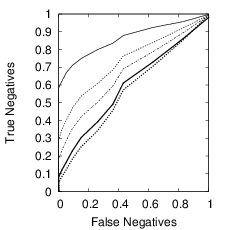}&
\includegraphics[width=0.2\linewidth]{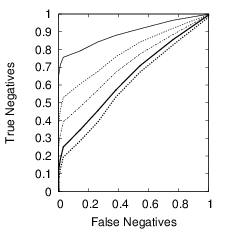}\\
Sample Size 20000&Sample Size 40000
\end{tabular}
\includegraphics*[width=0.8\linewidth]{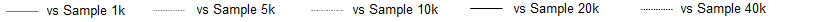}
\end{center}
\caption{ROC Curves for Five Sample Sizes: Query 3, Sampling Cheater Only}
\end{figure*}

\textbf{Cheater Sample Size} Figure 7 shows the ROC curves for query 3, against cheaters using sampling only.  We can clearly see that the curves move down and to the right as our adversary's sample size increases.  This makes sense, as the cheater gets better at impersonating the correct result, it becomes more difficult to distinguish the incorrect results from the correct ones.  However, in every case, even with a sample size of 1000, we are able to detect cheating better than random guessing.  Keep in mind that, in order to be useful, we merely need to be able to detect cheating with some non-negligible probability, and that any means we choose to do that is acceptable.

\begin{figure*}
\begin{center}
\begin{tabular}{ccc}
\includegraphics[width=0.2\linewidth]{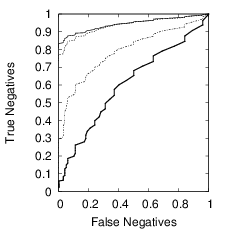}&
\includegraphics[width=0.2\linewidth]{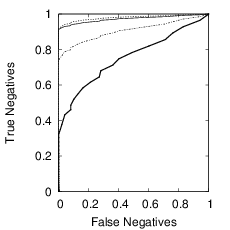}&
\includegraphics[width=0.2\linewidth]{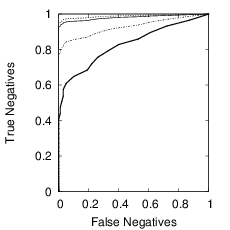}\\
Sample Size 1000&Sample Size 5000&Sample Size 10000
\end{tabular}
\begin{tabular}{cc}
\includegraphics[width=0.2\linewidth]{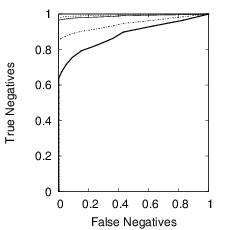}&
\includegraphics[width=0.2\linewidth]{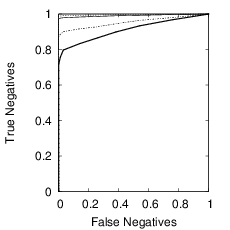}\\
Sample Size 20000&Sample Size 40000
\end{tabular}
\includegraphics*[width=0.8\linewidth]{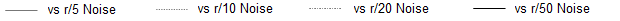}
\end{center}
\caption{ROC Curves for Five Sample Sizes: Query 3, Noisy Cheater Only}
\end{figure*}

\textbf{Cheater Laplace Noise}  Figure 8 shows the same query as figure 7, but this time with our cheater only using the Laplace noise.  Surprisingly, the noise addition version ends up being easier to catch.  This is due to the fact that the parameter on the Laplace noise is large enough to cause issues.  Still, at $r$/50, the cheating becomes quite difficult to detect for low sample size.

\section{Conclusions}

In summary, by thinking about the problem of query verification from a different perspective, namely, that of an economist, we can drastically reduce the computation required to ensure that the result you asked for is the result you received.  The various query verification methods that are out there are still quite useful, however.  Specialized verification methods which take up very little space work well for common queries, and in our game-theoretic framework, would only be required to run a fraction of the time.  They are, however, not generic and can rely on some expensive operations.  No matter what sort of verification method we choose, our contract-based computation vastly simplifies the overall process of query verification.

\subsection{Future Work}

In the future, we will consider other types of verification methods, and how they might be better served by the game-theoretic framework outlined here.  In addition, we will consider joins, which are disproportionately resource intensive compared to other database operations, possibly resulting in a need for a revised incentive structure.

\section{Acknowledgements}

This work was partially supported by Air Force Office of Scientific Research MURI Grant FA9550-08-1-0265, National Institutes of Health Grant 1R01LM009989, National Science Foundation (NSF) Grant Career-0845803, and NSF Grant 0964350.

\bibliography{gametheoryqueryverification}

\end{document}